\documentclass{WileyMSP-template}
\usepackage{amsmath}
\usepackage{amssymb}
\usepackage{bm}
\usepackage{color}
\usepackage{siunitx}

\usepackage[draft=false]{hyperref}

\usepackage[version=4]{mhchem}
\usepackage{upgreek}

\usepackage{xr}
 \definecolor{blueviolet}{rgb}{0.54,0.17,0.89}
\def\ket#1{\left| #1 \right\rangle}

\newcommand{\abs}[1]{\left| #1 \right|}

  \usepackage{color}

\def\vec#1{{\bf#1}}

\def\be{\begin{equation}}
\def\ee{\end{equation}}
\def\bea{\begin{eqnarray}}
\def\eea{\end{eqnarray}}

\begin{document}
\pagestyle{fancy}
\rhead{\includegraphics[width=2.5cm]{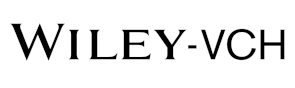}}

\title{Dark and Bright Excitons in Halide Perovskite Nanoplatelets}

\maketitle

\author{Moritz Gramlich\footnote[1]{Contributed equally}}
\author{Michael W. Swift\footnotemark[1]$^{,}$\footnote[2]{ASEE Postdoctoral Fellow}}
\author{Carola Lampe}
\author{John L. Lyons}
\author{Markus D{\"o}blinger}
\author{Alexander L. Efros*}
\author{Peter C. Sercel*}
\author{Alexander S. Urban*}

\begin{affiliations}
Moritz Gramlich, Carola Lampe, Prof. Dr. Alexander S. Urban\\
Nanospectroscopy Group, Nano-Institute Munich, Department of Physics, Ludwig-Maximilians-Universit{\"a}t (LMU), Munich 80539, Germany\\
Email Address: urban@lmu.de\\

Dr. Markus D{\"o}blinger\\
Department of Chemistry, Ludwig-Maximilians-Universit{\"a}t (LMU) \& Center for NanoScience (CeNS), 81377 Munich, Germany\\

Dr. Michael W. Swift, Dr. John L. Lyons, Dr. Alexander L. Efros\\
Center for Computational Materials Science, U.S. Naval Research Laboratory, Washington D.C. 20375, United States\\
Email Address: sasha.efros@nrl.navy.mil\\

Dr. Peter C. Sercel\\
Department of Applied Physics and Materials Science, California Institute of Technology, Pasadena, California 91125, United States\\
Center for Hybrid Organic Inorganic Semiconductors for Energy, Golden, Colorado 80401, United States\\
Email Address: psercel@caltech.edu\\
\end{affiliations}

\keywords{Halide Perovskite, Nanoplatelets, Quantum Confinement, Exciton Fine Structure, Photoluminescence Spectroscopy,  Effective Mass Model, Optoelectronics}


\begin{abstract}
Semiconductor nanoplatelets (NPLs), with their large exciton binding energy, narrow photoluminescence (PL), and absence of dielectric screening for photons emitted normal to the NPL surface, could be expected to become the fastest luminophores amongst all colloidal nanostructures. However, super-fast emission is suppressed by a dark (optically passive) exciton ground state, substantially split from a higher-lying bright (optically active) state. Here, the exciton fine structure in 2-8 monolayer (ML) thick \ce{Cs_{n-1}Pb_nBr_{3n+1}} NPLs is revealed by merging temperature-resolved PL spectra and time-resolved PL decay with an effective mass modeling taking quantum confinement and dielectric confinement anisotropy into account. This approach exposes a thickness-dependent bright-dark exciton splitting reaching \SI{32.3}{meV} for the 2ML NPLs. The model also reveals a 5-16 meV splitting of the bright exciton states with transition dipoles polarized parallel and perpendicular to the NPL surfaces, the order of which is reversed for the thinnest NPLs, as confirmed by TR-PL measurements. Accordingly, the individual bright states must be taken into account, while the dark exciton state strongly affects the optical properties of the thinnest NPLs even at room temperature. Significantly, the derived model can be generalized for any isotropically or anisotropically confined nanostructure. 
\end{abstract}

\section{Introduction}

The growing interest in colloidal nanoplatelets (NPLs),  which are atomically flat, quasi-two-dimensional sheets of semiconductors,  is justified by their potential  to become the best luminophores amongst all colloidal nanostructures.  The  lateral size of the typically rectangular-shaped NPLs is much larger than their thickness, 
which can be varied in a precisely controlled manner from two to almost a dozen monolayers (MLs). \cite{2017_Riedinger_NatMat}  Resulting NPL samples  are remarkably uniform and nearly monodisperse in the pre-determined thickness.   The growth of collodial quantum wells, such as the NPLs or nanoribbons was first reported  for II-VI compounds (e.g., CdSe, CdTe, CdS) \cite{2008_Ithurria_JACS,2011_Ithurria_NatMat,2006_Joo_JACS} and subsequently for lead halide perovskites. \cite{2015_Tyagi_JPCL,2015_Bekenstein_JACS,2016_Akkerman_JACS,2016_Shamsi_JACS,2020_Huo_NanoLett,2021_Liu_ACSnano} Optical investigations show that  these structures  have  electronic properties similar to freestanding quantum wells. Accordingly, the band edge absorption exhibits excitonic  transitions between the lowest 2D subbands of electrons and  holes, incrementally shifting to higher energies with decreasing thickness.  Due to the degeneracy of the valence band in II-VI compound NPLs,  one observes two absorption peaks associated with the lowest subbands of light and heavy holes in these systems,\cite{2011_Ithurria_NatMat,2020_Ji_ACSnano} similar to absorption spectra of epitaxially grown quantum wells.  In perovskite NPLs with non-degenerate valence and conduction bands, the absorption shows only one band-edge excitonic transition. \cite{2015_Bekenstein_JACS,2016_Weidman_ACSnano}  In both cases,  the photoluminescence (PL)  comes from the  lowest absorption subband and  does not display a significant Stokes shift.

Four essential factors influence the outstanding luminescence properties of NPLs: (i) An immense exciton binding energy ($E_X$ =100-500 meV) results from the {strong} spatial confinement of carriers in the direction perpendicular to the NPL surface. \cite{EfrosBrus2021} The electron-hole Coulomb interaction is significantly enhanced due to the small dielectric  constant of the media surrounding the NPLs, which reduces the dielectric screening. Consequently,  the effective radius of the 2D exciton, $a^*$, is tiny, and thus the exciton radiative decay rate, proportional to  $1/(a^*)^2$, is strongly increased. (ii) The second factor enhancing the radiative decay rate is the narrow emission linewidth, which results from the  low inhomogeneous broadening of the NPL samples due to an absence of a thickness fluctuation of the NPLs. (iii) There is a   strong coupling of photons to NPL excitons whose transition dipoles {are} parallel {to} the NPL surface. 
 This is because dielectric screening does not reduce the magnitude of the photon electric field for polarization parallel to the NPL surface. (iv) Finally, at low temperatures, the exciton giant oscillator strength connected with the coherent exciton motion in NPLs leads to a further increase of the radiative  recombination rate. \cite{1987_Feldmann_PhysRevLett} In  CdSe NPLs, the shortening of the radiative decay time \cite{2011_Ithurria_NatMat,2012_Tessier_ACSnano} and giant  oscillator transition strength~\cite{2021_Geiregat_Light,Naeem2015}
 have  been reported already, suggesting that all the above-discussed phenomena could bring the radiative decay time of  various  NPLs down to the range of tens to a few hundreds of ps.

A severe  obstacle  potentially prohibiting the realization of such super-fast emission from NPLs results from the fine structure splitting.
The electron-hole exchange interaction is known to split the  exciton into a lower energy dark (optically passive) level  split from   higher-lying bright (optically active)  levels. \cite{1996_Norris_PhysRevB,1996_Efros_PhysRevB,2017_Sercel_NanoLett,2004_Shabaev_NanoLett,2018_Shornikova_Nanoscale} 
A result of this level ordering is that at very low temperatures and assuming a thermalized population distribution only the dark exciton state is occupied, and the NPL decay time increases up to microseconds. However, in NPLs with a bright-dark splitting $\Delta E_{BD}$ far smaller than the thermal energy at room temperature $E_{th}\approx\SI{26}{meV}$, all states are nearly equally populated after excitation and so, even at room temperature, the dark exciton state decreases the radiative decay  rate of the bright exciton, $1/\tau_r$,  by  a factor equal to the ratio of the number of the bright exciton states to the number of total exciton states. For halide perovskites this amounts to $k_r=\frac{3}{4}\tau_r^{-1}$ and for II-VI compounds to $k_r=\frac{1}{2}\tau_r^{-1}$.
Exciton fine-structure splitting controlled by  the electron-hole  exchange interaction  is enhanced in nanostructures  due to spatial  and dielectric confinement. Accordingly, in CdSe NPLs $\Delta E_{BD}$ was measured  to be thickness-dependent and to vary from 2  to  6 meV. \cite{2018_Shornikova_Nanoscale} The splitting decreases the decay rate at room temperature by a factor of 1/2 because the excitons in the dark state are thermally excited into the bright exciton state for $T>70 K$.  This is in contrast to CdSe  quantum dots, where $\Delta E_{BD}$ is on the order of 20 meV, and depopulation of the bright exciton state affects the radiative decay time  even  at room temperature. \cite{2003_Crooker_ApplPhysLett} 
   
The surprising decrease of the radiative decay time  with temperature previously observed in lead halide perovskite nanocrystals (NCs) 
led to the suggestion that  these nanostructures  
exhibit a bright-dark exciton inversion on the order of 1 meV due to the Rashba effect. \cite{Becker,2021_Liu_ACSnano}
Still, the exact origin and nature of the Rashba effect remain unclear.  Within the Rashba hypothesis,  one can use the fine structure splitting of $\sim1$meV to estimate the magnitude of the Rashba terms, which amount to 0.3-0.45\,eV\AA.
The Rashba effect can only occur in the absence of inversion symmetry. In NPLs,  this symmetry breaking  could be induced  by surface reconstructions or by the surface ligands  used to passivate the NPLs. \cite{2021_Jana}
Studies of \ce{CsPbBr3} NPLs show, however, that the dark-bright splitting of the exciton is surprisingly large, $\sim \SI{22}{meV}$ in 2ML thick NPLs. \cite{2020_Rossi_NanoLett} This is significantly larger than the Rashba-induced effective exchange~\cite{SercelJCP2019,SercelNL2019} and thus would likely negate the Rashba inversion if this were present. 
Accordingly, it is vital to precisely determine and verify the magnitude of the exchange splitting dependent on the anisotropy of the NPLs both theoretically and experimentally.

In this article, we show that PL spectra and  time-resolved (TR-)PL decays of 2-8 ML thick \ce{Cs_{n-1}Pb_nBr_{3n+1}} NPLs  in the 4-100 K temperature range are in excellent agreement with  effective mass modeling of the 2D exciton fine structure
taking into account only the electron-hole exchange interaction.
Consequently, these results do not necessitate the inclusion of a Rashba effect for their interpretation. We find a thickness-dependent bright-dark exciton splitting $\Delta E_{BD}$ reaching up to \SI{32.3}{meV} for the 2ML thick NPLs, constituting the largest ever reported value. \cite{2020_Rossi_JChemPhys,2020_Rossi_NanoLett} Our effective mass model successfully describes  the experimental  thickness dependence using the long-  and short-range (LR and SR) exchange interaction. Taking into account the anisotropy of the band edge Bloch functions induced by strong, one-dimensional confinement in the NPLs, we find the spatial and dielectric confinement to significantly enhance the interaction. Moreover, the model reveals a large splitting of the bright exciton states into a component with the transition dipole polarized in the plane of the NPLs and a component with the transition dipole oriented perpendicular to the NPL surface. Interestingly, we { predict} an inversion of the bright levels for the thinnest (2ML) NPLs, where the out-of-plane dipole becomes energetically favorable. 

The bright exciton splitting is between 5 and \SI{16}{meV} for the 2-8 ML NPLs, indicating that the bright levels cannot be approximated
 as a single state. We confirm { this finding} through TR-PL measurements, which show that a two-level bright-dark exciton model cannot accurately reproduce the observed lifetime dependence on temperature and NPL thickness.  Using the splittings calculated from the effective mass model, we employ a novel three-level model with two bright states split into in-plane and out-of-plane components 
energetically 
above a lower-lying dark state. We obtain
 { good} model agreement 
with the measured
 PL decay rates for all thicknesses and temperatures. Importantly, our findings and model are not only valid for halide perovskites but can be easily 
generalized
to other semiconductor systems that are well described by effective mass theory. Accordingly, the model should provide a valuable template for other excitonic systems, especially those exhibiting strong quantum confinement.

\section{Results}

\begin{figure}
\includegraphics{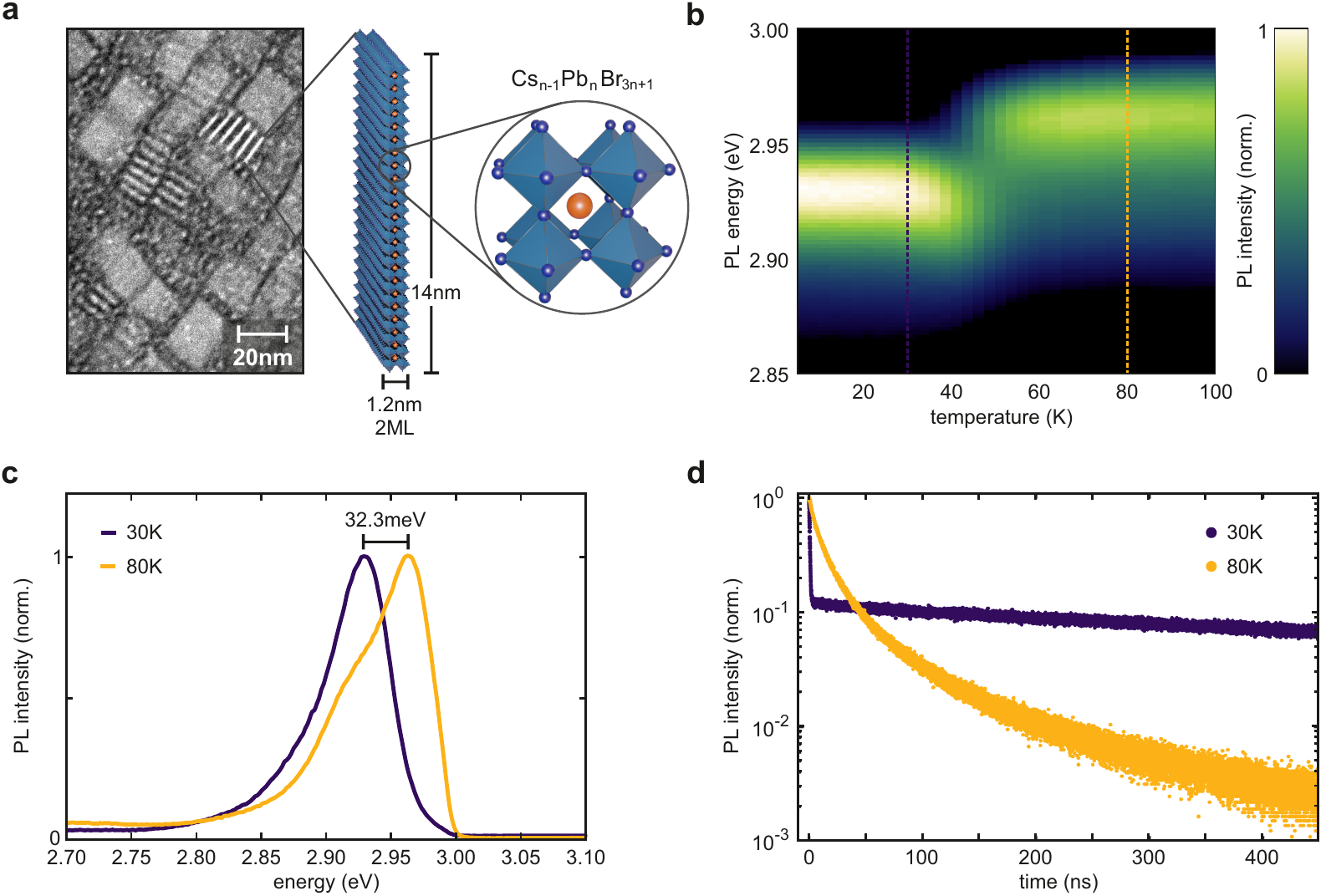}
\caption{\textbf{Temperature-dependent and time-resolved PL spectroscopy on 2ML CsPb\textsubscript{2}Br\textsubscript{7} NPL thin films.} $(\bf a )$ TEM image of the 2ML sample and a scheme depicting the structure of a 2ML NPL and its atomic crystal structure.  $(\bf b )$ Normalised, temperature-dependent PL spectra of a 2ML NPL thin film. The emission maximum undergoes a sharp \SI{32.3}{meV} large blueshift at \SI{50}{K}.  $(\bf c )$ Line traces taken from panel $(\bf b )$ at \SI{30}{K} (dark) and \SI{80}{K} (yellow) show the blueshift prominently. The spectrum at \SI{80}{K} appears to comprise two merged peaks. $(\bf d )$ Time-resolved PL decay at the respective emission maxima acquired at \SI{30}{K} and \SI{80}{K}. The low temperature decay exhibits two vastly different lifetimes, a fast one of $\tau_\mathrm{fast}<\SI{600}{ps}$ (faster than our IRF) and a slow component with $\tau_\mathrm{slow}>\SI{1}{\micro \second}$ indicative of a bright exciton level above a dark exciton level.}
\label{fig:1}
\end{figure}

The perovskite NCs used for this study were two-dimensional NPLs with the chemical composition \\
\ce{Cs_{n-1}Pb_nBr_{3n+1}}. The thickness of these NPLs assumes discrete values, which we denote as the number of monolayers, $n$,  of the NPL. Each ML 
 corresponds to
 the height of a \ce{[PbBr6]^{4-}} octahedron, approximately $5.9\ \text{\AA}$ (Figure \ref{fig:1}a). \cite{2016_Tong_AngChInt} Our synthesis of the NPLs is based on a slightly modified version of a previously reported synthesis \cite{2018_Bohn_NanoLett} (see Methods).
The resulting NPLs not only have higher quantum yields than previously reported, reaching up to 80\%, they are also extremely stable. In dispersion, they remain unchanged for a few weeks and even deposited onto substrates without further encapsulation, they remain stable for at least a week. This allows us to carry out detailed morphological characterization and spectroscopy on thin films and even individual NPLs. 
The as-synthesized NPLs have between two and eight MLs with corresponding thickness ranging from 1.2 to 4.7 nm. Importantly, each synthesis yields almost exclusively NPLs of only one thickness, as determined by linear optical spectroscopy. The NPLs are square-shaped with lateral dimensions of $14\pm2~$nm, as observed by scanning transmission electron microscopy in high-angle annular dark-field mode (STEM-HAADF) and shown here for the case of 2 ML NPLs (Figure \ref{fig:1}a, see Supporting Figure \ref*{sfig:TEM} and Supporting Table~\ref*{stab:dimensions} for other thicknesses). Oleylamine and oleic acid ligands passivate the NPLs to enhance their optical properties and stabilize them against fusion with other NPLs. As previously observed, the organic ligand shell leads to a minimum spacing of \SI{2.5}{nm} between the perovskite layers, which can be seen particularly in thin films. \cite{2020_Singldinger_ACSEnergy}
The strong confinement of the NPLs in one dimension leads to a significant shift of the absorption onsets and PL emission peaks from the bulk value of 2.36 eV \cite{2018_Lin_Nat} up to 2.85 eV for the 2MLs when dispersed in hexane (see Supporting Figure \ref*{sfig:ABS/PL}). 

Using a home-built $\upmu$-PL setup with excitation provided by a ps-pulsed white light laser and operating at temperatures down to \SI{4}{\kelvin}, we can investigate the temperature dependence of the NPLs' PL. For this, we drop-cast the NPL dispersions onto \ce{SiO2}-coated Si substrates. There is a notable shift of the PL as the temperature is decreased from room temperature to \SI{4}{\kelvin} due to thermal lattice expansion and exciton-phonon coupling. \cite{2019_Wang_JPCL} Interestingly, in the 2ML sample, there is a sharp jump of \SI{32.3}{\meV} in the emission wavelength at \SI{50}{K}, from \SI{2.96}{eV} down to \SI{2.93}{eV} (Figure  \ref{fig:1}b). The jump is also present in the spectra of NPLs of other thicknesses. However, the size of the jump and the temperature at which it occurs decrease with increasing NPL thickness, $d$ (see Supporting Figure \ref*{sfig:T-dependent_PL}).
Comparing the spectra at temperatures above and below this jump, we conclude that the emission stems from two separate energetic states rather than a gradual shift (Figure  \ref{fig:1}c). \ce{CsPbBr3}, both in bulk and in cube-shaped NCs, assumes its lowest energy orthorhombic phase already at room temperature. \cite{2020_Lopez_ACSOmega} Assuming this also to be the case for the NPLs, a phase-change cannot explain the observed jumps. Temperature-dependent TR-PL helps to shed light on the nature of these two states (Figure  \ref{fig:1}d). At temperatures above the jump, the PL decay can be reproduced well with a single exponential and a lifetime on the order of 10 ns. This lifetime becomes longer as the temperature is reduced. At a certain low temperature, the decay becomes bimodal with a short component below the instrument response function of our system of \SI{600}{ps} and a very long component on the order of \SI{1}{\micro\second}. The amplitude of the fast component increases rapidly as the temperature is further reduced and then remains constant for the lowest temperatures. These results indicate that the observed PL properties are a result of bright and dark exciton state emission. \cite{2018_Shornikova_Nanoscale,2020_Rossi_NanoLett} Accordingly, the PL jump observed above is due to a splitting of the dark and bright exciton: $\Delta E_{BD}=32.3$ meV. This constitutes the largest such splitting in any semiconductor system reported to date, being nearly six times as large as in CdSe NPLs of comparable thickness. \cite{2018_Shornikova_Nanoscale} This further supports the theoretical work stating that any possible energy level inversion induced by a Rashba effect would be negated by an enhanced electron-hole exchange interaction in strongly confined perovskite NCs. \cite{SercelNL2019,2020_Rossi_JChemPhys}

A two-level kinetic model, with a dark exciton and a slightly higher-lying bright exciton level, is typically employed to describe the excited states of the system. \cite{2020_Rossi_NanoLett} Excitons in each excited state can relax down to the ground state, or they can transition between the two excited states.  Applying this model to the TR-PL data, we find that the fit can reproduce the observed lifetimes. However, the fine-structure splitting values obtained do not match the experimentally deduced values (see Supporting Figure \ref*{sfig:2-level_model}). The bright exciton level is a triplet state, which is degenerate in the case of cubic crystal structure and isotropic crystal shape. \cite{SercelNL2019,Becker,2019_Tamarat_NatMat} In \ce{CsPbBr3} NCs with orthorhombic crystal structure 
and non-cubic shape, the degeneracy is lifted and the bright exciton states split in energy. The level splitting
was typically observed to be less than 2 meV.
 In NPLs, however, the shape anisotropy leads to a much larger splitting between the bright excitons with in-plane ($X$ and $Y$) and out-of-plane transition dipoles ($Z$), analogous to the aforementioned increased bright-dark exciton splitting. Inevitably, the bright triplet state can no longer be approximated by a single state. This in turn must induce a more complex behavior of the exciton dynamics.


\subsection*{Two-Dimensional  Exciton in Perovskite Nanoplatelets}
\begin{figure}
\includegraphics{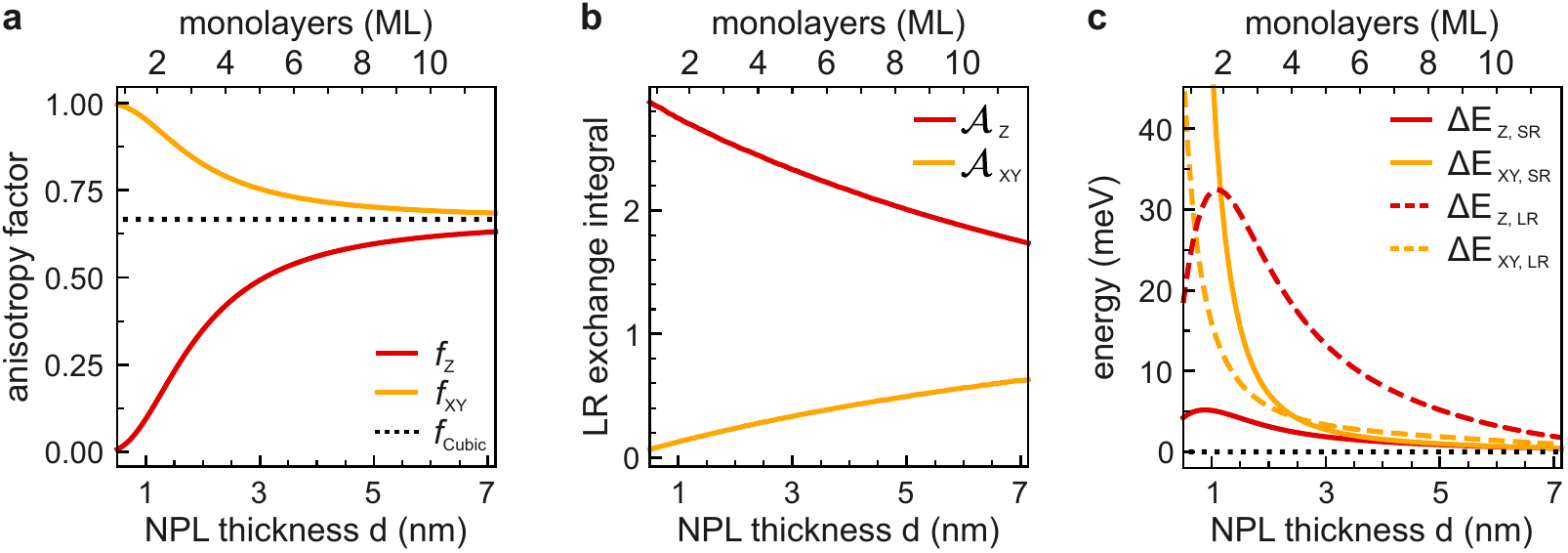}
\caption{\textbf{Important quantities calculated within the theoretical effective-mass model as a function of the NPL thickness.}  {$(\bf a )$} Fine structure anisotropy factors $f_{X_i}$ due to the anisotropic confinement of band edge Bloch functions in a square-shaped NPL.  Out-of-plane $f_Z$ is shown in red and in-plane $f_{XY}$ is in yellow. The value for an isotropic, cubic nanocrystal is 2/3 as depicted by the dashed black line. {$(\bf b )$} Dimensionless LR exchange integrals given by Equation~\eqref{eq:AZ'}. {$(\bf c )$} LR (dashed lines) and SR (solid lines) exchange contributions to the splitting between the dark exciton and out-of-plane bright exciton $Z$ (red) and between the dark exciton and in-plane bright exciton $XY$ (yellow).}
\label{fig:2}
\end{figure}

To explain the experimentally observed, unusually large bright-dark splitting and bright sublevel splitting, we have developed the effective-mass theory of excitons in 2D perovskite NPLs. As discussed in the introduction, both spatial confinement and dielectric confinement enhance the electron-hole Coulomb interaction and accordingly both the SR
and LR electron-hole exchange interaction.  The effective exciton  radius in these structures is much smaller than the lateral size of the NPLs. This allows for a separation of variables into independent  center-of-mass $\bm R= (m_e \bm r_e + m_h \bm r_h)/(m_e + m_h)$ and relative $\vec{r} = \vec{r}_e-\vec{r}_h$  coordinates. \cite{EfrosSovPhysSemicond86,SercelNL2019,SercelNS2020} The Hamiltonian is divided into two terms accordingly:
\be
\hat{H}_0 = \hat{H}_{0,{\rm com}} + \hat{H}_{0,{\rm rel}}={\hat {\bm P}^2\over 2M}+ \left[{\hat {\bm p}^2\over 2\mu}+ V(\bm r)\right]~,
\label{eqHamSeparated}
\ee
where we employ the exciton translational mass $M=m_e+m_h$, the reduced mass $\mu=(1/m_e+1/m_h)^{-1}$ as well as the center-of-mass momentum $\hat {\bm P}=\hat {\bm p}_e+\hat {\bm p}_h$ and relative momentum $\hat{\bm p} = (m_h\hat{\bm p}_e- m_e \hat{\bm p}_h)/M$. The Coulomb potential $V(\bm r) $ describing the electron-hole interaction takes into account the difference in dielectric constants of the semiconductor NPL and the surrounding media \cite{Ritova67,Keldish79} and is described using the Hanamura potential (see Supporting Methods~\ref*{methods:variational}). \cite{Hanamura,Hong}  The wavefunction of the exciton ground state in an infinite 2D semiconductor layer with thickness $d$ in the $z$-direction is made up of the band edge Bloch functions $u_{j_e}(\bm r_e)$ and $u_{j_h}(\bm r_h)$, the wavefunctions $\sqrt{2/d}  \cos(\pi z_e/d)$ and $\sqrt{2/d}  \cos(\pi z_h/d)$ that describe electron and hole confinement in the layer 
respectively,
and the relative motion wavefunction $\phi(\bm r_e - \bm r_h)$ with zero angular momentum in the $z$ direction. The wavefunction $\phi(\bm r_e - \bm r_h)$ cannot be found analytically, so we obtain it variationally using a 2D ``hydrogenic'' ansatz wavefunction:\cite{YangPRA}
\be
\phi_d(\bm r ) =  \frac{4}{a^*(d)} \frac{1}{\sqrt{2 \pi}} \exp\left(-{2|{\bm r}|\over a^*(d)}\right)~,
\label{eqAnsatz}
\ee
where $a^*(d)$, an effective  Bohr radius of the 2D exciton,  
is found using the variational procedure and  is strongly thickness dependent (Figure \ref{fig:2}a). 
The determined binding energies  (see Supporting Figure~\ref*{sfig:variational_phi}) match experimentally determined values for a variety of thicknesses. \cite{2018_Bohn_NanoLett,2019_Li_JPCL,2018_Blancon_NatComm}

The finite size of the NPL affects the exciton center-of-mass motion. Assuming that the edges of a NPL are impenetrable barriers for the exciton, we can write the exciton center-of-mass wavefunction as:\\  $2/\sqrt{L_xL_y}\cos(\pi X/L_x)\cos(\pi Y/L_y)$ for rectangular NPLs with edge lengths  $L_x$ and $L_y$. As a result the total exciton wavefunction in a NPL can be expressed as: 
\be
\Psi_{ j_e,j_h}^d(\bm r_e,\bm r_h) = 
\frac{4 u_{j_e}(\bm r_e) u_{j_h}(\bm r_h)  }{d\sqrt{L_xL_y}} \cos\left({\pi X\over L_x}\right) \cos\left({\pi Y\over L_y}\right)   \cos\left({\pi z_e\over d}\right)   \cos\left({\pi z_h\over d}\right)\phi_d(\bm r_e - \bm r_h)\ .
\label{eqExcitonWFNP}
\ee

\subsection*{Bloch function anisotropy}
\label{SecSlabModel}
In Equation \eqref{eqExcitonWFNP}, strong spatial confinement in the NPL significantly modifies the conduction band Bloch functions, $u_{j_e}(\bm r_e)$, compared to their bulk form 
for   cubic symmetry
shown in Ref. \cite{Becker}. Assuming that the perovskite layer of the NPL with an underlying cubic crystal symmetry is surrounded by an infinite confinement potential, we use a 6-band Luttinger Hamiltonian to describe the band edge Bloch functions, as detailed in the Methods section. The spatial confinement splits the 
 upper $J=3/2$ fourfold degenerate heavy/light electron states into two subbands similar to the effect of a tetragonal  crystal field. We show that the anisotropic part of the Hamiltonian can be expressed as:

\begin{align}
 \tilde{H}_2^S(0)= \left(
\begin{array}{cccccc}
 \Delta_{SO}+\frac{\delta }{3}  & 0 & 0 & 0 & 0 & 0 \\
 0 & \Delta_{SO} -\frac{\delta }{3} & 0 & 0 & \frac{\sqrt{2} \delta }{3} & 0 \\
 0 & 0 & \Delta_{SO} -\frac{\delta }{3} & 0 & 0 &-\frac{\sqrt{2} \delta }{3}  \\
 0 & 0 & 0 & \Delta_{SO}+\frac{\delta }{3}  & 0 & 0 \\
 0 & \frac{\sqrt{2} \delta }{3} & 0 & 0 & 0 & 0 \\
 0 & 0 &-\frac{\sqrt{2} \delta }{3} & 0 & 0 & 0 \\
\end{array}
\right)~,
\end{align}
where $\Delta_{SO}$ is the spin-orbit splitting between the $J=3/2$ (light electron, heavy electron) and the $J=1/2$ states (split-off electron) in the conduction band. The parameter $\delta$ can be considered as an effective
tetragonal
 crystal field and is given by:
\begin{align}
\delta = - 3 \frac{\hbar^2}{2m_0} \gamma _2 \frac{2 \pi ^2}{d^2}~,
\label{eq:delta}
\end{align}
where $\gamma_2 = E_p / 6E_g^\prime$ is the second Luttinger parameter,  $E_g^\prime$ is the band gap of the NPL (see Supporting Methods~\ref*{methods:slab}), and $E_p$ is the Kane energy parameter (see Table~\ref{tab:params}). These parameters induce a splitting of the conduction band into three bands, analogously to the splitting of the valence band in traditional semiconductors such as GaAs. The lowest-lying conduction band, also referred to as the split-off band, has an energy of:\cite{SercelNL2019}
\begin{align}
E_c = \frac{3\Delta_{SO} -\delta}{6} - \frac{1}{2}\sqrt{\Delta_{SO}^2 - \frac{2}{3} \Delta_{SO} \hspace{0.05in}\delta + \delta^2}
\label{eq:EnergiesTET}
\end{align}
and the resulting Bloch functions can be written as:\cite{SercelNL2019,Nagamune}
\begin{eqnarray}
u_{c,1/2} =- \sin\theta Z \uparrow - \cos\theta \frac{X + i Y }{\sqrt{2}}\downarrow
\, , & \qquad &
u_{c,-1/2} = - \cos\theta\frac{X - i Y}{\sqrt{2}} \uparrow + \sin\theta Z \downarrow \, .
\label{eqCBBlochTet}
\end{eqnarray}
In these expressions, the angle $\theta$ is given by:\cite{Nagamune}
\begin{align}
\tan 2\theta = \frac{2\sqrt{2} \Delta_{SO}}{\Delta_{SO} - 3 \delta}~, \hspace{0.3in} (\theta\le\frac{\pi}{2})~.
\label{eq:XstalfieldTheta}
\end{align}


\subsection*{Fine structure of the ground exciton state}

The LR and SR exchange interaction between electron and hole spin states split the fourfold degenerate ground exciton levels in perovskite NPLs described by the wavefunction in Equation \eqref{eqExcitonWFNP} into three optically active bright exciton states and one optically passive dark exciton state. The exciton fine structure is typically calculated by splitting the interaction into SR and LR components: $\Delta E_{exchange} = \Delta E_{SR} + \Delta E_{LR}$. We shall initially consider the SR exchange interaction.

\subsubsection*{Short-range (SR) electron-hole exchange interaction}
The short-range exchange interaction has the form of a contact interaction with an effective spin operator acting on the electron and hole Bloch functions:\cite{RosslerTrebin,SercelNL2019}
\begin{align}
\hat H_{\rm exch}^{SR} = {1\over 2}C^{SR} \Omega\left[ \mathbb{I}-(\bm \sigma_e \cdot\bm \sigma_h)\right]\delta(\bm r_e-\bm r_h)~.
\label{eqSRex}
\end{align}
In this expression, $C^{SR}$ is the exchange constant, $\Omega$ is the volume of the crystal unit cell, $\mathbb{I}$ is the 4x4 unit matrix and $\bm\sigma_{e,h}$ are the Pauli operators acting on the electron and hole spins.

To find the dark-bright exciton splitting one can directly rewrite Equation \eqref{eqSRex} in the matrix representation using four pairs of electron and hole Bloch functions and diagonalize the matrix. The Bloch functions of the conduction band $u_{c,\pm 1/2}$ are defined  in Equation \eqref{eqCBBlochTet} and  the Bloch functions of the $s$-like valence band are $u_{v,1/2} = |S\uparrow\rangle$ and $u_{v,-1/2}  = |S\downarrow\rangle$. It is more convenient, however,  to transform the pair Bloch function basis to the basis of 
linear
 dipoles. \cite{SercelNS2020} Herein, the basis functions $|X_i\rangle$ run over $|D\rangle$,$|X\rangle$,$|Y\rangle$,$|Z\rangle$, which represent the wavefunctions of the dark state and the three bright excitons with transition dipoles aligned along the $X,Y,Z$ directions. \cite{SercelNS2020} 

As we show in Supporting Methods~\ref*{methods:SR}, the SR exchange Hamiltonian in this basis can be expressed as:
\begin{align}
\tilde{H}_{\mathrm{exch}}^{SR} =  C^{SR}  \Theta\left(
\begin{array}{cccc}
 0 & 0 & 0 & 0 \\
 0 & \cos^2\theta &0 & 0 \\
 0 & 0 &\cos^2\theta& 0 \\
 0 & 0 & 0 & 2\sin^2\theta \\
\end{array}
\right)~,
\end{align}
where $\theta$ is defined in Equation \eqref{eq:XstalfieldTheta} and the overlap factor $\Theta$ represents the probability that the electron and hole reside in the same unit cell. \cite{SercelNL2019}  For the quasi-2D exciton with envelope  wavefunction $f(\bm{r}_e,\bm{r}_h)=\frac{2}{\sqrt{L_xL_y}} \cos(\frac{\pi X}{L_x})\cos(\frac{\pi Y}{L_y})\frac{2}{d}\cos(\frac{\pi z_e}{d}) \cos(\frac{\pi z_h}{d})\phi_d(\bm r_e - \bm r_h)$  (see Equation \eqref{eqExcitonWFNP} ) this factor is given by the expression:
\begin{align}
\Theta=   \Omega \iint_V d^3r_ed^3r_h  f^\ast(\bm{r}_e,\bm{r}_h) \delta(\bm{r}_e-\bm{r}_h) f(\bm{r}_e,\bm{r}_h)=\frac{3}{2d}\Omega |\phi_d(0)|^2~. 
\label{eq:Theta2dQW}
\end{align}

The splitting introduced by the SR exchange interaction between the dark exciton and bright excitons with two degenerate $X$ and $Y$ dipoles can be written accordingly:
\be
\Delta E_{X, SR} = \Delta E_{Y, SR}= f_X C^{SR} \Omega \frac{3}{2d} \abs{\phi_d(0)}^2~,~
\label{eq:13}
\ee
and between the dark exciton and the bright $Z$ dipole:
\be
\Delta E_{Z, SR} =  f_Z C^{SR} \Omega \frac{3}{2d} \abs{\phi_{d}(0)}^2 ~,~
\label{eq:14}
\ee
where $f_Z = 2\sin^2(\theta)$ and $f_X =f_Y= \cos^2(\theta)$.  The difference between Equation \eqref{eq:13} and \eqref{eq:14} describes the splitting between the bright excitons in the square NPLs for the states with dipoles in-plane $(X,Y)$ and out-of-plane $(Z)$.


\subsubsection*{Long-range (LR) electron-hole exchange interaction}
Next, we can consider the LR exchange interaction, which in bulk semiconductors induces a splitting between the longitudinal and transverse (LT) optically active exciton states. To calculate the LR exchange interaction in a NPL, we use the straightforward formalism developed by Cho. \cite{ChoJPSJ} Cho's  analysis begins with the observation that any   exciton state is accompanied by a transition dipole density $\bm{\mathcal P}(\vec{r})$, which is induced by a charge density defined as  $\rho(\vec{r}) = - \bm{\nabla}\cdot \bm{\mathcal P}(\vec{r})$. Cho showed that in this case the LR exchange can be expressed in terms of the Coulomb interaction of this  charge  density  associated with a given  exciton state:\cite{ChoJPSJ}
 \begin{align}
H_{X_i}^{LR}= \int_{V_1} dV_1 \int_{V_2} dV_2 \left[ -\bm \nabla_1\cdot \bm{\mathcal P}_{X_i}(\bm r_1) \right]^\ast \frac{1}{\epsilon_\infty |\bm r_1 -\bm r_2|} \left[ -\bm \nabla_2\cdot \bm{\mathcal P}_{X_i}(\bm r_2) \right]~,
\label{eqCho}
\end{align}
where  $\epsilon_\infty$ is the high frequency dielectric constant that screens  the Coulomb interaction. 

Using the linear dipole exciton Bloch function basis  we identify the  polarization $ \bm{\mathcal P}$  
as  the transition dipole moment density associated with 
a given
exciton state $X_i$,  whose wavefunction is described as $f(\bm r_e,\bm r_h)   \ket{X_i}$.  For such a transition the polarization can be written as:
\be
 \bm{ \mathcal P}_{X_i}(\bm r_e)=i{\hbar e\over m_0 \hbar\omega}\int d^3r_h \hspace{0.05in} f(\bm r_e,\bm r_h)\hspace{0.05in} {\bm p}_{X_i}\hspace{0.05in} \delta(\bm r_e-\bm r_h) = i{\hbar e\over m_0 \hbar \omega}  f(\bm r_e,\bm r_e)\hspace{0.05in} {\bm p}_{X_i}~,
\label{eqPolarizDefn}
\ee
where we define the transition energy $E_2-E_1=\hbar \omega$ (see Supporting Figure~\ref*{sfig:hbar_omega_PL})  and $ {\bm p}_{X_i}$ is the  matrix element of the momentum operator taken between the Bloch functions of the conduction and valence bands  for the  $X_i$  exciton state. Previously, we successfully used this procedure to calculate the LR exchange interaction in quasicubic perovskite nanocrystals \cite{SercelNL2019,SercelJCP2019} and perovskite nanowires \cite{FolieJPCA2020}.
In  Supporting Methods~\ref*{methods:LR}, we 
apply
the same procedure 
to
 square-shaped NPLs, with 
dimensions  $L\times L\times d$.  
Straightforward calculations yield  the following expression for the dark-bright exciton splitting for the exciton states  $X_i$:
\be
\Delta E_{X_i,LR}(d)
= \frac{\hbar \omega_{LT}(d)}{2}f_{X_i}  \mathcal A_{X_i}(d/L)  \left(\frac{\Theta}{\Theta_\text{bulk}}\right)~,
\label{eqShapeH}
\ee
where $\hbar\omega_{LT}(d) = \hbar\omega_{LT}^\text{bulk} (E_g/\hbar\omega)^2$ is the $d$-dependent LT splitting energy. The bulk exchange overlap $\Theta_\text{bulk} = \Omega/(\pi a_x^3)$ is determined by the bulk exciton radius $a_x$.  $\mathcal A_{X_i}(d/L)$ are dimensionless integrals defined in Equation \eqref{eq:AZ'} that depend on the ratio of the NPL thickness to their lateral size, $d/L$.  The dependence of  $\mathcal A_{X}$ and  $\mathcal A_{Z}$ on the NPL thickness is calculated numerically 
and shown in Figure \ref{fig:2}b and Supporting Figure \ref*{sfig:LR}.  One can see that as $d/L \to 0$,   $\mathcal A_{X}\rightarrow 0$ and $\mathcal A_{Z}\rightarrow 3$.  In square-shaped NPLs the $XY$ dipole states remain degenerate, while the LR exchange interaction splits these states for the case $L_x\neq L_y$.

Combining the SR and LR exchange interaction, we obtain the total splitting of the dark state to the bright $Z$ and $(X,Y)$ states in a square NPL:
\bea
    \Delta E_Z &= &{3 f_Z\over 2d}\abs{\phi_d(0)}^2\left[ C^{SR} \Omega   + \frac{\hbar\omega_{LT}(d)}{2} \pi a_x^3  \mathcal   A_Z(d/L)\right]~,\nonumber\\
    \Delta E_X &= & \Delta E_Y={3 f_X\over 2d}\abs{\phi_d(0)}^2\left[ C^{SR} \Omega   + \frac{\hbar\omega_{LT}(d)}{2} \pi a_x^3  \mathcal   A_X(d/L)\right]~.
\eea

\begin{figure}
\includegraphics[width=\textwidth]{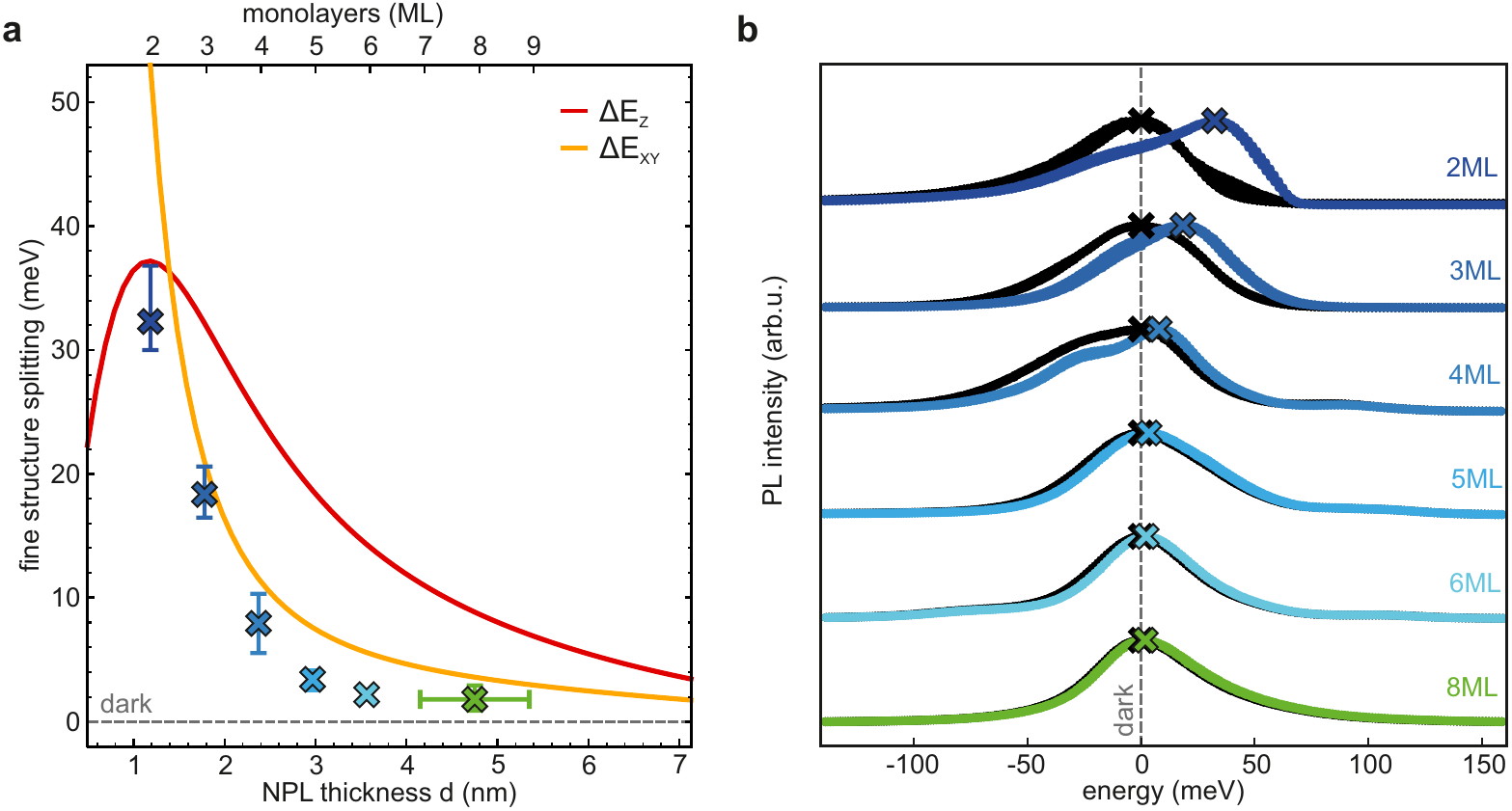}
\caption{\textbf{Thickness dependence of the bright-dark exciton level splitting in NPLs.} $(\bf a )$ Total exchange splitting (SR and LR) for the bright out-of-plane $Z$ (red line) and in-plane $(X,Y)$ (orange line) exciton states from the dark state in a square NPL with lateral side length $L=14$ nm. The dark exciton level is the lowest for all thicknesses. The bright  $Z$ level lies above the $XY$ level for all thicknesses above the 2ML NPLs. Splitting energies determined experimentally from the shift of the PL spectra, as depicted in $(\bf b )$. Horizontal error bars on the 8ML sample indicate a possible minor contribution from 7ML and 9ML NPLs. The y-error bars denote the standard deviation of the PL maxima analyzed in $(\bf b )$ to obtain the energetic position of the exciton levels. $(\bf b )$ Temperature-dependent PL spectra of NPL thin films. For each thickness, several spectra below the jump (black dots) and above the jump (colored dots) are superimposed to help determine the location of the dark and lowest bright exciton states, denoted by the crosses. Temperature intervals in which the PL jump occurs are shown in Supporting Table 2. All spectra are offset horizontally according to the respective energetic position of the dark exciton level of each NPL. Accordingly, the PL jump becomes larger as the thickness of the NPL decreases, from \SI{5}{meV} for 8MLs up to \SI{32.3}{meV} for 2MLs.
}
\label{fig:3}
\end{figure}

The calculated dependence of these energies on \ce{Cs_{n-1}Pb_nBr_{3n+1}} NPL thickness is plotted in Figure \ref{fig:3}a (solid lines), where the dark exciton level is set to $\Delta E=0$.  The
material
 parameters used for this calculation are given in Table~\ref{tab:params}. Importantly, the bright excitons are always located energetically above the dark exciton state. The previously reported bright-dark level inversion in \ce{CsPbBr_3} NCs was due to the Rashba effect. \cite{Becker}  However, the magnitude of the exchange interaction and the crystal structure anisotropy in \ce{CsPbBr_3} NCs are significantly smaller than the corresponding effects  in the NPLs. Hence, for the NPLs the addition of possible Rashba terms would not affect the ordering of dark and bright excitons and are not considered here.  For nearly all NPL thicknesses, the in-plane bright states $X,Y$ are energetically lower than the out-of-plane bright states $Z$, as one would intuitively expect.  Interestingly, there is a crossing of these states at $d\leq1.4$ nm, below which the exciton state with out-of-plane transition dipole is the energetically favorable one. This $d$ value lies between the 2ML and 3ML NPL thickness. Accordingly, the polarization of PL emitted from the lowest bright exciton state should be parallel to the NPL surface for all NPLs except for the thinnest 2ML NPLs, for which the PL should be polarized perpendicular to the NPL surface. Furthermore, the calculations show that the total splitting between ($X,Y$) and $Z$ bright excitons exhibits a prominent thickness dependence. Increasing initially from 8 meV to 13.2 meV from 2 to 4MLs, the splitting decreases to 4.5 meV for the 8ML NPLs. Accordingly, one cannot assume the bright states to act as one degenerate level, revealing  why the two-level model could not explain the TR-PL data.

\begin{table}
\begin{tabular}{llll}
\hline
Parameter & Value  & Description& Source\\
\hline
$\hbar\omega_{LT}^\text{bulk}$ & $5.4\ \mathrm{meV}$     & LT exciton splitting & Experiment \cite{Bayer19}\\
$E_p$              & $27.8\ \mathrm{eV}$     & Kane energy & Derived from $\hbar\omega_{LT}$ \cite{Bayer19,SercelNL2019}\\
$a_x$              & $3.1\ \mathrm{nm}$      & bulk exciton Bohr radius& Theory \cite{SercelJCP2019}\\
$\epsilon_i$       & 7.3                     & interior dielectric constant& Experiment \cite{Yang}\\
$\epsilon_o$       & 2.1263                  & exterior dielectric constant & Experiment  \cite{CRC2020}\\
$\epsilon_\infty$  & 4.76                    & high-frequency dielectric constant & Derived from  $\hbar\omega_{LT}^\text{bulk}$ and $E_p$ \cite{Bayer19,SercelJCP2019}\\
$\mu$              & $0.126\, m_0$           & reduced effective mass & Experiment \cite{Yang} \\
$m_e$              & $0.25\, m_0$            & electron effective mass & Theory \cite{SercelJCP2019}\\
$m_h$              & $0.25\, m_0$            & hole effective mass & Theory \cite{SercelJCP2019}\\
$E_g$              & $2.342\ \mathrm{eV}$    & band gap & Experiment ($T=4.2$ K) \cite{Yang}\\
$\Delta_{SO}$      & $1.5\ \mathrm{eV}$      & spin-orbit splitting & Theory \cite{SercelNL2019}\\
$\Omega$           & $0.2104\ \mathrm{nm}^3$ & pseudocubic unit-cell volume& Experiment \cite{SercelNL2019}\\
$C^{SR}$           & $315.5 \ \mathrm{meV}$  & short-range exchange constant& Derived from experiment \cite{Aich20,SercelNL2019} \\
\hline
\end{tabular}
\caption{\textbf{\ce{CsPbBr3} material parameters used for calculation of the level splitting.}  Parameters are taken from the literature as indicated.  The ligands bonded to the nanocrystal surface make up the exterior dielectric, so we take $\epsilon_o$ to be the square of the refractive index of oleic acid.  The short-range exchange constant $C^{SR}$ is calculated based on experiments in \ce{FAPbBr3}~\cite{Aich20} together with $\Omega$ and $a_x$, using the methodology in Ref.~\cite{SercelNL2019}.  Calculated values of $C^{SR}$ in \ce{CsPbBr3} and \ce{FAPbBr3} are similar~\cite{SercelNL2019} but larger than the experimental value for \ce{FAPbBr3}, so we use the experimental value in this work.}
\label{tab:params}
\end{table}


\subsection*{Spectroscopically determined bright-dark splitting}
 
Having determined the thickness-dependent bright-dark level splitting for the NPLs, we can revisit the PL spectra of the ensembles (Figure \ref{fig:3}b). As mentioned earlier, all samples exhibit a jump in the PL maximum at a given temperature. The jump is very prominent in the 2 ML sample and can easily also be observed in the 3 ML and 4 ML samples. The small shoulder on the low energy side of the 4 ML spectra is due to a small sample inhomogeneity, likely due to varying lateral sizes. This however does not affect the analysis here. For the thicker NPLs, the PL jump becomes more of a gradual shift and is more difficult to discern by eye. To quantify the jump, we must first consider the gradual PL shift arising from the combination of exciton-phonon coupling and lattice expansion.\cite{ 2019_Wang_JPCL} These effects are strongly thickness and temperature-dependent. We determine their contribution to the overall PL shift for each thickness in the temperature ranges around the jump (see Supporting Table \ref{Tab:Temperatures}). The total shift of approximately 2 meV is rather low (see Figure \ref{fig:1} and Supporting Figure \ref{sfig:T-dependent_PL}), consequently, for all but the 8ML NPLs it can be completely neglected. Accordingly, by plotting several spectra just below (black dots) and just above the jump (colored dots), we can more easily determine the spectral positions of the bright and dark states. Notably, for each thickness, the temperature ranges are in good agreement with the transition temperatures predicted by the extracted energetic width of the jumps and with the PL decay curves discussed below. The crosses in Figure 3b denote the respective PL maxima, and the spectra are offset according to the spectral position of the dark state. It is evident that the shift between the two maxima, which corresponds to $\Delta E_{BD}$, increases with diminishing thickness of the NPLs from 1.8 meV for the 8 ML NPLs to 32.3 meV for the 2 ML NPLs. Accordingly, the PL jump can only be easily discerned for the thinner NPLs and appears as an increased shift for the thicker NPLs. 
Importantly, experimental and theoretical splittings are in excellent agreement, especially for the thinnest NPLs. Interestingly, the value for the 2 ML sample lies closer to the curve for the $Z$ dipole, supporting the notion of a bright exciton level inversion for the thinnest NPLs. Polarization-dependent spectroscopy on single NPLs could fully confirm this prediction. For the thicker NPLs, the experimental values lie below the predicted ones. This could be due to the fact that determining the splitting becomes progressively more difficult as the splitting is reduced. Also, at low temperatures with a reduced linewidth broadening of the PL spectra, one can see that they are not fully monodisperse and contain a small population of NPLs of different thickness. TEM images also confirm this (see Supporting Figure \ref*{sfig:TEM}). The slight polydispersity could affect the measurements for the thicker NPLs, where the PL positions do not vary much between NPLs with $\Delta \mathrm{ML} = \pm1$.  Another aspect that we have so far disregarded is the fine structure splitting resulting from the crystal structure anisotropy itself. The NPLs assume the orthorhombic crystal structure, which should lead to an additional energy level splitting. \cite{2019_Bertolotti_ACSnano,2020_Huo_NanoLett} While this is small compared to $\Delta E_{BD}$ in the thinner NPLs; the same cannot be said for the thicker NPLs. If the crystal structure aligns preferentially along one NPL direction, this could either reduce or increase the overall splitting of the exciton fine structure.


\subsection*{Exciton lifetimes}

\begin{figure}
\includegraphics[width=\textwidth]{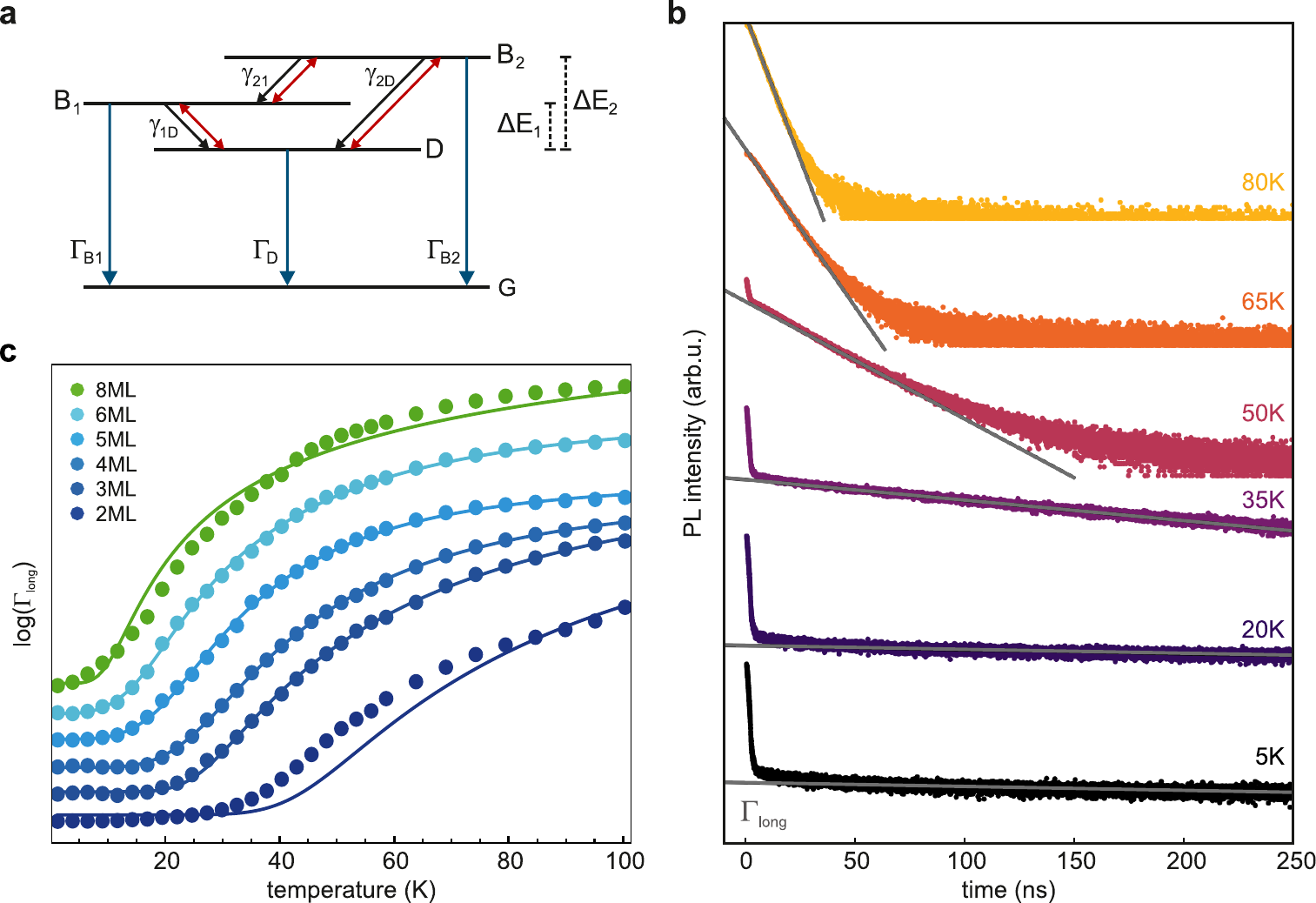}
\caption{\textbf{PL decay in the NPLs explained by a three-level model.} $(\bf a )$ Three-level model developed to explain the time-resolved PL data. Three excited states (a lowest-lying dark state, D, and two bright states, $B_1$ and $B_2$) can decay directly to the ground-state, G. Excitons can also exchange between the three excited states via one-phonon interaction. Bright-dark splitting between the two bright levels is given by $\Delta E_{1,2}$. $(\bf b)$ Series of temperature-dependent PL decay traces between \SI{5}{K} and \SI{80}{K} for 4ML NPL thin films. The decay rate of the slow component, $\Gamma_\mathrm{long}$, which is determined from an exponential fit function, becomes progressively faster as the temperature is increased. $(\bf c)$ The slow decay rate $\Gamma_\mathrm{long}$ in dependence of temperature for all NPL samples (colored dots, with color coding corresponding to Figure~\ref{fig:3}). The data points are plotted logarithmically (in ns$^{-1}$) and offset vertically for clarity. All samples show the same trend with a constant decay rate, which rapidly becomes larger starting from a given temperature. This temperature, coinciding with the experimentally observed jump in the PL spectra, decreases with increasing NPL thickness. This is due to a reduced exciton level splitting and an easier thermal activation from lower lying levels. A fit of $\Gamma_\mathrm{long}$ employing the three-level model (solid lines) shows good agreement to the experimental data for all thicknesses.}
\label{fig:4}
\end{figure}

The PL decay times could not be explained with the commonly used two-excited-level model \cite{2020_Rossi_NanoLett}  as mentioned above. As we showed in the last section, the bright level splitting is between 5-16 meV for the NPLs studied here. Consequently, one should instead consider a three-level system (see Figure \ref{fig:4}a). In this model, the lowest excited state D lies beneath two bright levels denoted $B_1$ and $B_2$, as the $X,Z$ ordering changes between 2ML and 3ML. Optical excitation occurs far above these levels, hence, excitons will relax quickly, populating all three excited levels. At low temperatures, the bright states will be emptied rapidly, as their recombination rates, $\Gamma_{B_1}$ and $\Gamma_{B_2}$ are very large. Excitons in the dark state reside there for a long time, since the direct recombination rate $\Gamma_{D}$ is very low. Non-radiative recombination processes are not included in this model;  such processes would act to reduce the overall quantum yield if they were significant. With increasing temperature, a thermally induced (phonon-mediated) transition between the excited states becomes more likely. Thus, excitons in the dark state can also decay indirectly via thermal excitation into either of the bright states. Accordingly, at the lowest temperatures, the prolonged decay observed in the TR-PL measurements (denoted $\Gamma_\mathrm{long}$) corresponds exactly to $\Gamma_{D}$. Thus, the rate can be obtained by fitting the long tail (Figure \ref{fig:4}b). At higher temperatures, other pathways are accessible, significantly complicating the population evolution of the dark state (see Supporting Methods~\ref*{methods:3-level_model}). Consequently, the slow decay observed $\Gamma_\mathrm{long}$  is no longer equal to $\Gamma_{D}$. Nevertheless, we can model this decay rate and compare the calculations to the TR-PL data using the three-excited-level model. To reduce the number of free parameters for fitting, we have fixed the energy levels of the three states and the ratio $\Gamma_{B_1}/\Gamma_{B_2}$ according to the results obtained from the theoretical model described above. The slow decay rate $\Gamma_\mathrm{long}$ was measured for all NPL thicknesses and is plotted for the temperature range 4-100 K (Figure \ref{fig:4}c, filled circles; the curves are offset vertically for clarity). At the lowest temperatures, $\Gamma_\mathrm{long}$ is constant, but then rapidly increases. The temperature for which this change occurs is strongly dependent on the NPL thickness. While the decay rate remains roughly constant up to 25 K for the 2ML sample, for the 8ML sample, there is nearly no constant rate. 
These results agree with the previously determined thickness-dependent bright-dark exciton splitting, as thermal excitation from the dark state is activated at lower temperatures for increasing NPL thickness. The fits resulting from the 3-level model (solid lines) accurately reproduce the experimental data for all NPLs, especially for the 3-6 ML NPLs, validating the model. The less-than-optimal fitting in the case of the 8ML sample is likely due to the inhomogeneity of the sample, resulting in different lifetimes and also due to energy transfer via F\"{o}rster resonance energy transfer occurring. \cite{2020_Singldinger_ACSEnergy} In the 2ML sample, the small mismatch in the range 30-80 K is caused by two effects. First, the emergence of a self-trapped exciton in this temperature range (see Supporting Figure~\ref*{sfig:self-trapped}) means there is an additional decay pathway from excitons residing in the dark state. \cite{2020_Paritmongkol_JPCL} The other effect is the effective-mass model's tendency to slightly overestimate  $\Delta E_{1,2}$. 
As the model still has five free parameters, the fit is qualitative rather than quantitative. Yet, it demonstrates that the calculated splitting energies can explain the TR-PL data. Additional experiments, including the temperature dependence of the fast PL decay rate, are necessary to improve the model to extract meaningful transition rates.

\section{Discussion}

This study demonstrates and quantifies the influence of shape anisotropy on the exciton fine structure in 2D halide perovskite NPLs. For the first time, we show that this anisotropy leads to an energetic splitting of the bright triplet into in-plane and out-of-plane levels. Temperature-dependent PL-spectroscopy on 2-8 ML thick NPLs reveals that in all NPLs independent of their thickness the lowest exciton sublevel is an optically forbidden dark exciton. The bright-dark exciton splitting increases from \SI{1.8}{meV} for the 8ML up to \SI{32.3}{meV} for the 2ML NPLs, constituting the largest value ever measured. Using an effective-mass model, we calculate the exciton fine structure in the NPLs.  The model quantitatively explains this large splitting, showing it is completely described by the electron-hole exchange interaction. This interaction is significantly enhanced by the strong confinement of carriers in the out-of-plane direction in the NPLs. Notably, the magnitude of these shifts shows that the bright levels cannot be considered degenerate and must be individually accounted for. Accordingly, we can use the splitting energies and the decay rates obtained from the temperature-dependent TR-PL measurements to calculate the expected PL spectra for each NPL sample (see Supporting Figure \ref{sfig:theory_PL}). These PL maps show striking agreement with the experimentally obtained ones (compare Supporting Figure \ref{sfig:T-dependent_PL}) with the slight exception for the 2ML case. There, an non-negligible emission from the bright-states at 0K is likely due to the aforementioned overestimation of the bright-bright level splitting. Nevertheless, the agreement between the calculations and experiment strongly supports our interpretation of the data and the model of the excitonic fine structure we have developed here.   

Interestingly, the model predicts an inversion of the bright exciton states for NPLs thinner than 3ML. Accordingly, for all NPLs but the 2MLs, the polarization of the lowest optically allowed transition lies in the NPL plane, while in the 2MLs it  is oriented perpendicular to the NPL plane. Polarization-dependent spectroscopic studies of single NPLs are needed to confirm this prediction.  Using the energy level splitting determined from the model, we can reproduce and explain the temperature dependence of the TR-PL data for all NPL samples. This fitting supports the three-level model of the 2D system with a lowest-lying dark exciton and two bright exciton levels split by several meV.  With splitting energies of this magnitude, an impact should still be visible at room temperature, especially for the thinnest NPLs. In two previous studies we had observed that the PL decayed faster in thicker than in thinner NPL. \cite{2018_Bohn_NanoLett,2020_Singldinger_ACSEnergy} With a concomitant increase in quantum yield and a reduced nonradiative decay rate, this meant that a larger radiative decay rate must be the cause of the faster PL decay in the thicker NPLs. However, this stood in contrast to the behavior of quantum wells, for which it was deduced that the radiative decay rate should increase as the quantum well is reduced in size. \cite{1987_Feldmann_PhysRevLett} The findings in this work can help to explain the apparent contradiction if one considers that even at room temperature an exciton will spend a considerable amount of time in the lowest dark state. Thermal excitation to a bright state (feeding) will lead to a prolonged PL emission from this bright state, masking its actual radiative decay rate. 

To describe the experimental results in \ce{Cs_{n-1}Pb_nBr_{3n+1}}  NPLs  we have developed a model that takes into account the anisotropy of the band edge Bloch functions introduced by spatial confinement. Additionally, we establish a tractable and straightforward approach to the calculation of the LR exchange interaction. These effects are essential in our system, and their inclusion will likely be necessary for other systems exhibiting strong shape anisotropy. The model can easily be extended to rectangular NPLs 
\\($L_X \neq L_Y$, both $\gg d$) by simply replacing $L$ in the LR exchange integrals as appropriate.  The model is also broadly generalizable to any semiconductor NPL system that is well described by effective mass theory.  We believe it will be a valuable template for use in other excitonic systems such as in multilayered 2D lead halide perovskites. \cite{2020_Paritmongkol_JPCL}

Further refinement to the model could be provided by incorporating both shape and crystal structure anisotropy. For this, one would need to know whether and how the crystal structure aligns with the \\ nanocrystal anisotropy. Polarization-dependent PL spectroscopy predominantly on single NPLs could help to elucidate this question. Nevertheless, our study provides a novel theoretical model to determine the energy level structure in highly anisotropic nanostructures, essential for optimizing these further and incorporating them into optoelectronic applications.

The level separation for the 2-8 ML NPLs is larger than the Rashba-induced effective exchange previously predicted in 3D \ce{CsPbBr3} NCs~\cite{SercelNL2019}, validating the choice to neglect the Rashba terms in our model, and suggesting that the ground exciton state in \ce{CsPbBr3} NPLs will remain dark in the absence of factors that substantially enhance the Rashba effect.  The search for NPLs with sufficiently strong Rashba interactions to exhibit a bright ground exciton should be continued.

\section{Methods Section}

\threesubsection{Synthesis}

Materials: \ce{Cs_2CO_3} (cesium carbonate, 99 \%), \ce{PbBr2} (lead(II) bromide, $\ge$ 98 \%), oleic acid (technical grade 90 \%), oleylamine (technical grade 70 \%), acetone (for HPLC, $\ge$ 99.9 \%), toluene (for HPLC, $\ge$ 99.9 \%) and hexane (for HPLC, $\ge$ 97.0 \%, GC) were purchased from Sigma-Aldrich. All chemicals were used as received.

The \ce{PbBr2} and Cs precursor solutions, as well as the PL enhancement solution, were prepared according to Bohn {\it et al.},~\cite{2018_Bohn_NanoLett} the synthesis of the NPLs, however, was  modified somewhat.

The synthesis was carried out under ambient atmosphere at room temperature. A vial was charged with \ce{PbBr2} precursor solution and Cs precursor solution was added under stirring at \SI{1200}{rpm}. Find the corresponding volumes for different thicknesses in Supporting Table~\ref*{SITable:2}. After \SI{10}{\second}, acetone (\SI{2}{\milli\litre}) was added and the reaction mixture was stirred for \SI{1}{\minute}. The mixture was centrifuged at \SI{4000}{rpm} for \SI{3}{\minute} and the precipitate was redispersed in hexane (\SI{1.8}{\milli\litre}). To enhance the stability and emission properties of the NPLs, an enhancement solution (\SI{200}{\micro\litre}) was added. 

\threesubsection{Electron Microscopy}

Scanning transmission electron microscopy in high-angle annular dark field (STEM-HAADF) mode was performed with a probe-corrected Titan Themis (FEI) at an acceleration voltage of 300 kV. 

\threesubsection{PL Spectroscopy}
\label{methods:PL}

Steady-state and time-resolved PL measurements were conducted using a pulsed laser (NKT Photonics, SuperK Fianium FIU-15) and an excitation wavelength of $\mathrm{\gamma_{central}=410\,nm}$. The laser was operated with a repetition rate of 1.95 MHz and an excitation power of $\mathrm{9\,Wcm^{-2}}$. The light was collected with a 50x objective (Mitutoyo, Plan Apo HR 50x) with a working distance of 5.2 mm. Steady-state PL spectra were measured using a SpectraPro HRS-500 spectrometer with a $\mathrm{300\,mm^{-1}}$ grating and a PIXIS charge-coupled device (all Teledyne Princeton Instruments). Time-resolved PL measurements were conducted using a SPAD (Excelitas Technologies, SPCM-AQRH-16-BR1) and a TCSPC-device (Swabian Instruments, Time Tagger 20). The silicon substrates (300 nm \ce{SiO2} layer) with the nanoplatelets drop-casted onto them were cooled using an attoDRY 800 closed-cycle liquid helium cryostat (Attocube). Room-temperature PL (and absorption) of the dispersions was measured using a commercial spectrometer \\(FLUOROMAX-Plus, HORIBA). 

\threesubsection{Bloch function anisotropy}

The anisotropy factors $f_z$ and $f_{XY}$ (see Figure~\ref{fig:2}a) reflect the effect of the anisotropy between the in-plane and out-of-plane directions in a NP on the band-edge Bloch functions.  The Bloch function of the valence band has $s$-symmetry and can  be  written as $u_{v,1/2}= | S\rangle| \uparrow \rangle$ and $u_{v,-1/2}=|S\rangle| \downarrow \rangle$,  where  the term  $|S\rangle$ denotes an orbital state  which transforms as the real spherical harmonic  with  $l=0$ while $ |\uparrow\rangle,$ $|\downarrow\rangle$ are the usual spin eigenstates. \cite{Becker}

For the conduction bands, the Bloch functions are more complex due to the $p$-symmetry of these bands.
The conduction bands of perovskite semiconductors can be described using the 6-band Luttinger Hamiltonian in the spherical approximation.
In a basis of Bloch functions $|J,J_z\rangle$ taken in the order $|3/2, 3/2\rangle$, $|3/2, 1/2\rangle$, $|3/2, -1/2\rangle$, $|3/2, 3/2\rangle$, $|1/2, 1/2\rangle$, and  $|1/2, -1/2\rangle$, the Hamiltonian has the form:\cite{1966_Pidgeon_PhysRev}
\be
\tilde{H} = \tilde{H}_1+\tilde{H}_2
\ee
where
\be
\tilde{H}_1 = \frac{1}{2m_0} \left(
\begin{array}{cccccc}
 \gamma _1 p^2 & 0 & 0 & 0 & 0 & 0 \\
 0 & \gamma _1 p^2 & 0 & 0 & 0 & 0 \\
 0 & 0 & \gamma _1 p^2 & 0 & 0 & 0 \\
 0 & 0 & 0 & \gamma _1 p^2 & 0 & 0 \\
 0 & 0 & 0 & 0 & \gamma _1 p^2 & 0 \\
 0 & 0 & 0 & 0 & 0 & \gamma _1 p^2 \\
\end{array}
\right)
\ee
with  $p^2=p_x^2+p_y^2+p_x^2$ and,

{\small\be
 \tilde{H}_2 = \frac{1}{2m_0}  \nonumber\left(
\begin{array}{cccccc}
 \Delta_{SO} +\gamma _2 p_\mathrm{eff}^2 & -2 \sqrt{3} \gamma _2 p_- p_z & -\sqrt{3} \gamma _2 p_-^2 & 0 & \sqrt{6} \gamma _2 p_- p_z & \sqrt{6} \gamma _2 p_-^2 \\
 -2 \sqrt{3} \gamma _2 p_+ p_z & \Delta_{SO} -\gamma _2 p_\mathrm{eff}^2 & 0 & -\sqrt{3} \gamma _2 p_-^2 & \sqrt{2} \gamma _2 p_\mathrm{eff}^2 & -3 \sqrt{2} \gamma _2 p_- p_z \\
 -\sqrt{3} \gamma _2 p_+^2 & 0 & \Delta_{SO} -\gamma _2 p_\mathrm{eff}^2 & 2 \sqrt{3} \gamma _2 p_- p_z & -3 \sqrt{2} \gamma _2 p_+ p_z & -\sqrt{2} \gamma _2 p_\mathrm{eff}^2 \\
 0 & -\sqrt{3} \gamma _2 p_+^2 & 2 \sqrt{3} \gamma _2 p_+ p_z & \Delta_{SO} +\gamma _2 p_\mathrm{eff}^2 & -\sqrt{6} \gamma _2 p_+^2 & \sqrt{6} \gamma _2 p_+ p_z \\
 \sqrt{6} \gamma _2 p_+ p_z & \sqrt{2} \gamma _2 p_\mathrm{eff}^2 & -3 \sqrt{2} \gamma _2 p_- p_z & -\sqrt{6} \gamma _2 p_-^2 & 0 & 0 \\
 \sqrt{6} \gamma _2 p_+^2 & -3 \sqrt{2} \gamma _2 p_+ p_z & -\sqrt{2} \gamma _2 p_\mathrm{eff}^2 & \sqrt{6} \gamma _2 p_- p_z & 0 & 0 \\
\end{array}
\right)\,
\ee}
where $p_\mathrm{eff}^2 = \left(p_x^2+p_y^2-2 p_z^2\right)$ and $\gamma_1$ and $\gamma_2$ are the Luttinger parameters.

The Hamiltonian $\tilde{H}_1$ is isotropic and while it affects the band dispersion it does not affect Bloch function mixing so we drop it from discussion and focus on $\tilde{H}_2$.  We note that the upper  conduction bands are four fold degenerate at $k=0$ and have $J=3/2$.  These we label    as heavy-electrons  with $J=3/2,$ $J_z = \pm 3/2,$ and light-electrons with $J=3/2,$ $J_z = \pm 1/2$.  The lower spin-orbit split off band has $J=1/2,$ $J_z = \pm 1/2,$  and is the lowest conduction band. 

The   Luttinger Hamiltonian above is expressed in the basis of Bloch functions at the band edges, namely the kets $ \ket{J,J_z}$;  in a bulk system the conduction band states will be described by Bloch waves of the form,
\begin{align}
\psi_{\bm k}(\bm r) =
\frac{e^{i(\bm k \cdot \bm r)} }{\sqrt{V}} \sum_{J,J_z} C_{J,J_z} \ket{J,J_z},
\end{align}
whose expansion coefficients are found by diagonalizing $H(\bm k)$ for a given wave vector.

The conduction bands in a perovskite slab are described by the Luttinger Hamiltonian above, but we are specifically interested in the form of solutions appropriate for  a slab geometry.  Given a slab of area $A$ assumed large, with   thickness $d$ centered at $z=0$, the requirement that the wavefunction vanish at the slab surface
 constrains the eigenstates  associated with the lowest slab subband in the slab   to have the   form,
\begin{align}
\psi_{1,\bm k}(\bm r) =\sqrt{\frac{2}{d}} \cos\left(\frac{\pi z}{d}\right)
\frac{e^{i(k_x x +k_y y)} }{\sqrt{A}} \sum_{J,J_z} C_{J,J_z} \ket{J,J_z}~,
\end{align}
where the $C_{J,J_z}$ again are expansion coefficients.  This expression motivates a change of basis from the Bloch function basis,$\ket{J,J_z}$, to what we will call the ``slab'' basis $ \ket{S} \ket{J,J_z}$, where $\langle z \ket{S} \equiv S(z) =\sqrt{\frac{2}{d}} \cos(\pi z/d)$.  
Transforming to this basis, the Hamiltonian $\hat{H}_{2}$ is represented by $\tilde{H}_{2}^{S} = \int_{d/2}^{d/2} dz S(z)\tilde{H}_{2} S(z) $ giving the following result:
{\small \begin{align}
\tilde{H}_2^S= \frac{\hbar^2}{2m_0}
\left(
\begin{array}{cccccc}
 \gamma _2 k_\mathrm{tot}^2+\Delta_{SO}  & 0 & -\sqrt{3} \gamma _2 k_-^2 & 0 & 0 & \sqrt{6} \gamma _2 k_-^2 \\
 0 & \Delta_{SO} -\gamma _2 k_\mathrm{tot}^2 & 0 & -\sqrt{3} \gamma _2 k_-^2 & \sqrt{2} \gamma _2 k_\mathrm{tot}^2 & 0 \\
 -\sqrt{3} \gamma _2 k_+^2 & 0 & \Delta_{SO} -\gamma _2 k_\mathrm{tot}^2 & 0 & 0 & -\sqrt{2} \gamma _2 k_\mathrm{tot}^2 \\
 0 & -\sqrt{3} \gamma _2 k_+^2 & 0 & \gamma _2 k_\mathrm{tot}^2+\Delta_{SO}  & -\sqrt{6} \gamma _2 k_+^2 & 0 \\
 0 & \sqrt{2} \gamma _2 k_\mathrm{tot}^2 & 0 & -\sqrt{6} \gamma _2 k_-^2 & 0 & 0 \\
 \sqrt{6} \gamma _2 k_+^2 & 0 & -\sqrt{2} \gamma _2 k_\mathrm{tot}^2 & 0 & 0 & 0 \\
\end{array}
\right),
\end{align}}
where $k_\mathrm{tot}^2 = \left(-\frac{2 \pi ^2}{d^2}+k_x^2+k_y^2\right)$.  All linear order terms involving $p_z$ vanish.  This slab Hamiltonian  can be broken into two pieces as follows:
\begin{align}
\tilde{H}_2^S = \tilde{H}_2^S(0) + \tilde{H}_2^S(\bm k)~,
\end{align}
where,
{\small \begin{align}
\tilde{H}_2^S(\bm k)= \frac{\hbar^2}{2m_0}
\left(\begin{array}{cccccc}
 \gamma _2 k_\perp^2 & 0 & -\sqrt{3} \gamma _2 k_-^2 & 0 & 0 & \sqrt{6} \gamma _2 k_-^2 \\
 0 & -\gamma _2 k_\perp^2 & 0 & -\sqrt{3} \gamma _2 k_-^2 & \sqrt{2} \gamma _2 k_\perp^2 & 0 \\
 -\sqrt{3} \gamma _2 k_+^2 & 0 & -\gamma _2 k_\perp^2 & 0 & 0 & -\sqrt{2} \gamma _2 k_\perp^2 \\
 0 & -\sqrt{3} \gamma _2 k_+^2 & 0 & \gamma _2 k_\perp^2 & -\sqrt{6} \gamma _2 k_+^2 & 0 \\
 0 & \sqrt{2} \gamma _2 k_\perp^2 & 0 & -\sqrt{6} \gamma _2 k_-^2 & 0 & 0 \\
 \sqrt{6} \gamma _2 k_+^2 & 0 & -\sqrt{2} \gamma _2 k_\perp^2 & 0 & 0 & 0 \\
\end{array}
\right)\ .
\end{align}}
Here, $k_\perp^2 = \left(k_x^2+k_y^2\right)$.  The more interesting part is $ \tilde{H}_2^S(0)$:
\begin{align}
 \tilde{H}_2^S(0)= \left(
\begin{array}{cccccc}
 \frac{\delta }{3}+\Delta_{SO}  & 0 & 0 & 0 & 0 & 0 \\
 0 & \Delta_{SO} -\frac{\delta }{3} & 0 & 0 & \frac{\sqrt{2} \delta }{3} & 0 \\
 0 & 0 & \Delta_{SO} -\frac{\delta }{3} & 0 & 0 & -\frac{\sqrt{2} \delta }{3} \\
 0 & 0 & 0 & \frac{\delta }{3}+\Delta_{SO}  & 0 & 0 \\
 0 & \frac{\sqrt{2} \delta }{3} & 0 & 0 & 0 & 0 \\
 0 & 0 & -\frac{\sqrt{2} \delta }{3}  & 0 & 0 & 0 \\
\end{array}
\right) ,
\end{align}
where the parameter $\delta$ can be considered as an effective crystal field, and is given by,
\begin{align}
\delta = - 3 \frac{\hbar^2}{2m_0} \gamma _2 \frac{2 \pi ^2}{d^2} ~.
\end{align}

\medskip
\textbf{Supporting Information} \par 
Supporting Information is available from the Wiley Online Library or from the author.

\medskip
\textbf{Acknowledgements} \par 
Al. L. E. and  J. L L. acknowledge support from the US Office  of Naval Research and the Laboratory-University Collaboration Initiative (LUCI) program of the DoD Basic Research Office; Theoretical calculations of exciton fine structure, long-range exchange interactions, and shape anisotropy effects were supported by the Center for Hybrid Organic Inorganic Semiconductors for Energy (CHOISE) an Energy Frontier Research Center funded by the Office of Basic Energy Sciences, Office of Science within the US Department of Energy; A.S.U. gratefully acknowledges support from the Bavarian State Ministry of Science, Research and Arts through the grant “Solar Technologies go Hybrid (SolTech)”, the Center for NanoScience (CeNS), the Deutsche Forschungsgemeinschaft (DFG) under Germany’s Excellence Strategy EXC 2089/1-390776260 and from the European Research Council Horizon 2020 through the ERC Grant Agreement PINNACLE (759744). M. W. S. acknowledges support from the Naval Research Laboratory Postdoctoral Fellowship through the American Society for Engineering Education.

\medskip

%



\section*{Conflict of Interest}
The authors declare no conflicts of interest.


\newpage

\begin{figure}
\textbf{Table of Contents}\\
\medskip
  \includegraphics{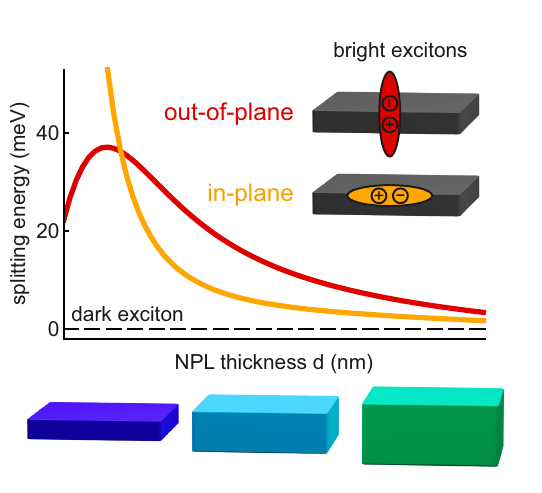}
  \medskip
  \caption*{Two-dimensional semiconductor nanoplatelets can enable faster and more efficient light-emitting devices. For this, a detailed understanding of their energetic structure is paramount. Here, the thickness-dependent exciton fine structure of nanoplatelets is deduced by merging temperature and time-resolved photoluminescence spectroscopy with a novel effective mass model, considering anisotropic quantum and dielectric confinement. Relevantly, the model can be generalized for any nanostructure.} 
\end{figure}

\end{document}